\documentclass[prd,aps,prd,twocolumn,showpacs,preprintnumbers,nofootinbib,amsmath,amssymb, superscriptaddress]{revtex4-1}

\usepackage{array}
\usepackage{graphicx}
\usepackage{bm}
\usepackage{amsmath, amsthm,amssymb,url}
\usepackage[normalem]{ulem}
\usepackage{color}
\usepackage{subcaption}
\usepackage{color}
\usepackage{graphicx}
\usepackage{float}
\usepackage{url}
\usepackage{xcolor}
\usepackage{listings}
\usepackage{array}

\DeclareMathOperator*{\argmax}{arg\,max}

\def\be{\begin{equation}}
\def\ee{\end{equation}}
\def\bea{\begin{eqnarray}}
\def\eea{\end{eqnarray}}
\def\bn{\begin{enumerate}}
\def\en{\end{enumerate}}

\def\tcb{\textcolor{blue}}

\makeatletter

\newcommand{\Rmnum}[1]{\expandafter\@slowromancap\romannumeral #1@}
\makeatother
\newcommand{\norm}[1]{\left\lVert#1\right\rVert}
\newcommand{\abs}[1]{\left|#1\right|}

\begin{document}
\captionsetup[figure]{labelfont={bf},labelformat={default},labelsep=period,name={Figure.}}
\preprint{------}

\title{
 Astrophysical signal consistency test adapted for gravitational-wave 
 transient searches 
 } 

\author{V.~Gayathri}
\affiliation{Department of Physics, Indian Institute of Technology Bombay, Powai, Mumbai, Maharashtra 400076, India}

\author{P.~Bacon}
\affiliation{APC, Univ Paris Diderot, CNRS/IN2P3, CEA/Irfu, Obs. de Paris, Sorbonne Paris Cit\'{e}, F-75013 Paris, France}

\author{A.~Pai}
\affiliation{Department of Physics, Indian Institute of Technology Bombay, Powai, Mumbai, Maharashtra 400076, India}

\author{E.~Chassande-Mottin}
\affiliation{APC, Univ Paris Diderot, CNRS/IN2P3, CEA/Irfu, Obs. de Paris, Sorbonne Paris Cit\'{e}, F-75013 Paris, France}

\author{F.~Salemi}
\affiliation{Max Planck Institute for Gravitational Physics (Albert Einstein Institute), D-30167 Hannover, Germany}

\author{G.~Vedovato}
\affiliation{INFN, Sezione di Padova, I-35131 Padova, Italy}

\date{\today}
\begin{abstract}
Gravitational wave astronomy is established with direct observation of gravitational wave from merging binary black holes and binary neutron stars during the first and second observing run of LIGO and Virgo detectors. The gravitational-wave transient searches mainly separate into two families: modeled and modeled-independent searches.  The modeled searches are based on matched filtering techniques and model-independent searches are based on the extraction of excess power from time-frequency representations. We have proposed a hybrid method, called  wavegraph that mixes the two approaches. It uses astrophysical information at the extraction stage of model-independent search using a mathematical graph.  In this work, we assess the performance of wavegraph clustering in real LIGO and Virgo noise (the sixth science run and the first observing run) and using the coherent WaveBurst transient search as a backbone. Further, we propose a new signal consistency test for this algorithm. This test uses the amplitude profile information to distinguish between the gravitational wave transients from the noisy glitches. This test is able to remove a large fraction of loud glitches, which thus results in additional overall sensitivity in the context of searches for binary black-hole mergers in the low-mass range.

\end{abstract}


\maketitle

\section{Introduction}
\label{SecI}
Gravitational-wave (GW) astronomy began with the first direct detection by the two Advanced LIGO detectors (Hanford and Livingston, US) \cite{TheLIGOScientific:2014jea} of the signal from binary black hole (BBH) merger on September 14th 2015 \cite{GW150914-DETECTION}. During their first
and second
observing runs, the LIGO detectors have further observed jointly with the Virgo detector \cite{TheVirgo:2014hva} (Pisa, Italy) ten BBH merger signals~\cite{GW150914-DETECTION,GW151226-DETECTION,GW170104-DETECTION,GW170608-DETECTION,GW170814-DETECTION,TheLIGOScientific:2016pea,O1-O2-catalog-paper} and one binary neutron star merger \cite{GW170817-DETECTION}. The third observing run started on April 1st, 2019. The LIGO and Virgo collaborations have, since, reported the detection of several compact binary candidates. During the next ten years the LIGO and Virgo detectors will continue their observations while gradually approaching their design sensitivity. KAGRA (Japan) \cite{Aso:2013eba} and a third LIGO instrument in India \cite{LIGO-India} are expected to join the network of observing detectors end of 2019 and around 2025, respectively.

Compact binary coalescences (CBC) composed of NS and/or stellar-mass BHs are primary sources for ground based detectors. The GW signal is buried in the detector noise. To detect the GW transient signals, different search methodologies have been developed, that can be categorized into two types: model-based and model-independent searches.

The model-based search, also referred to as matched filtering search aim to in find the best fitting waveform in a discrete set of physical model waveforms, called \textit{templates}. While this type of search is powerful, it has two limitations. It can be computationally expensive when the number of templates to compare with is large. In the second LIGO-Virgo observing run, approximately $400,000$ template waveforms are obtained assuming spin-aligned, non-precessing compact coalescing binaries ~\cite{Cannon:2011vi,Privitera:2013xza,Messick:2016aqy,Canton:2014ena,Usman:2015kfa,Authors:2019qbw}. These number of templates increases under more general assumptions. For instance, it increases by a factor of 10 when considering precessing binaries with arbitrarily aligned spins~\cite{harry16:_searc_gravit_compac, Indik:2016qky}. This type of search also relies on the availability of accurate waveform models. This is not true in all areas of the signal parameter space, in particular for precessing or eccentric binary systems \cite{harry16:_searc_gravit_compac,Bustillo:2015qty,Varma:2014jxa,Capano:2013raa,Huerta:2016rwp} where waveform development is still a field of active research.

Model-independent searches assume minimum prior information about the waveform morphology~\cite{Lynch:2015yin,Cornish:2014kda,Thrane:2015psa,Cornish:2014kda}. They are based on the extraction of excess power from time-frequency representations (TF maps) of the multi-detector observations. While computationally much cheaper, this type of search does not perform as \tcb{well} as the model-based search to detect CBC signals \cite{O1-O2-catalog-paper,O1-O2_imbhb_search}, primarily because they get affected by non-Gaussian and non-stationary features, also called ``glitches'', present in the detector noise \cite{Cabero:2019orq}.

In an earlier work~\cite{Bacon:2018zgb}, we proposed a hybrid method, called {\it wavegraph} that mixes the two approaches. This method is a search for time-frequency patterns defined from a range of astrophysical waveform models. The set of time-frequency patterns are mapped to a graph data structure, which allows to perform the pattern search very efficiently using combinatorial optimization algorithm. This method was implemented into the model-independent search algorithm Coherent WaveBurst (cWB) ~\cite{Klimenko:2008fu,klimenko05:_const,Klimenko:2015ypf}. 
We showed using simulated Gaussian noise that wavegraph improves the sensitive range of cWB to BBH signals by $5\%$ to $13\%$ depending on the total mass of the binary.

In this work, we assess the performance of wavegraph in real noise, including glitches. 
We propose a signal consistency test for this algorithm, similar to the glitch rejection techniques introduced in model-based searches, such as $\chi^2$ test \cite{Allen:2004gu,Nitz:2017lco,Canton:2013joa}, bank $\chi^2$ \cite{Dhurandhar:2017aan} or $\xi^2$ tests~\cite{Messick:2016aqy}. 
We demonstrate and validate this method in the context of BBH search using LIGO S6 and O1 data. 

In Sec. \ref{cWBsection}, we briefly summarize the coherent waveburst algorithm as well as the wavelet based, graph theoretical based clustering scheme developed in \ref{Sec:wavegraph}.  In Sec \ref{Sec:consistency_test}. we propose the signal consistency test for wavegraph clustering. In Sec. \ref{SecV} and \ref{SecVI}, we present the results of the simulation.

\section{Coherent WaveBurst and wavegraph algorithms}
\label{SecII}

In this section, we give a brief overview of the coherent WaveBurst and wavegraph algorithms. 

\subsection{Coherent WaveBurst algorithm}
\label{cWBsection}

The cWB algorithm is used to search for a broad class of weakly modelled GW transients in multi-detector data with minimum prior knowledge of the targetted signal. The algorithm has been successfully applied to eyes-wide-open, all-sky, and all-time searches for short-duration GW transients in the data from Advanced LIGO and Virgo observing runs \cite{Abbott:2016blz,TheLIGOScientific:2016uux,Aasi:2014iwa,Abadie:2012rq,Virgo:2012aa,Abadie:2010mt}. We now highlight the main features of the algorithm.

A time-frequency map is obtained using the Wilson-Daubechies-Meyer (WDM) transform \cite{necula12:_trans_wilson_daubec} that projects the detector data onto bases of functions localized in the time-frequency domain Multiple bases are considered with functions that span a range of durations defined by
the scale parameter $M$. We denote $w_k(t,f;M)$, the WDM transform data for the $k$-th detector in the multi-detector network.

The data for the whole network is collected in a vector ${\mathbf{w}_{\theta,\phi}[p]} \equiv \{ w_k(t-\tau_k(\theta,\phi),f;M)/\sqrt{S_k(f)} \}_{k=1\ldots K}$ where $S_k(f)$ is the noise power spectrum from the $k$-th detector and here we use the short-hand notation $p$ for the time, frequency and scale coordinates. This vector depends on the sky coordinates $\theta, \phi$ because we compensated for the propagation delay $\tau_k(\theta,\phi)$ between the detector $k$ and a reference, assuming the source is located in $(\theta, \phi)$. The pixel energy $E_{\theta,\phi}[p] = \norm{\mathbf{w}_{\theta,\phi}[p]}^2_2$ is maximized over $(\theta, \phi)$ and the collection of pixels with large maximum energy is retained \cite{Klimenko:2008fu,klimenko05:_const}.

Based on the salient pixels, the cWB algorithm forms clusters of neighboring pixels. Various clustering rules can be used based on the pixel geometrical vicinity in the time-frequency domain. The wavegraph method described in the next section follows a different approach based on the information inferred from a set of astrophysical waveform models.

The cWB algorithm then evaluates the significance of each of these clusters using a likelihood ratio under Gaussian noise assumptions. The log-likelihood ratio reads $L(\theta,\phi) = 2({\bf{w}}|{\boldsymbol{\xi}}) - ({\boldsymbol{\xi}}|{\boldsymbol{\xi}})$ where $\boldsymbol{\xi}$ is the noise-scaled network response to the impinging GW signal $h_+$ and $h_\times$ computed in the time-frequency domain. The components of this vector are $\xi_k[p] = [F_{k,+} h_+(t,f;M) + F_{k,\times} h_{\times}(t,f;M)]/\sqrt{S_k(f)}$ with $F_{k,+}(\theta,\phi)$ and $F_{k,\times}(\theta,\phi)$ are the antenna pattern function for the $+$ and $\times$ polarizations of the $k$-th detector. By maximizing over $h_+$ and $h_\times$, we obtain the statistic
\begin{equation}
\label{eq:maxLratio}
L_{max}(\theta, \phi) = \sum_{p \in C} {\bf{w}}_{\theta,\phi}[p]^T {\bf{P}}_{\theta,\phi}[p]  {\bf{w}}_{\theta,\phi}[p],
\end{equation}
which is used to select significant cluster as an event candidate.

This statistic is divided into two parts $L_{max} = E_{in} + E_{coh}$, where $E_{in}$ and $E_{coh}$ are the incoherent and coherent energies respectively which capture the diagonal and off-diagonal terms of the network projection operator ${\bf{P}}_{\theta,\phi}[p]$. The null energy 
$E_{null}=E_{tot}-L_{max}$ with $E_{tot}=\sum_{p \in C} E[p]$ captures the energy in the plane orthogonal to the network plane.

The network correlation coefficient \cite{Klimenko:2008fu} $c_c \equiv E_{coh}/(|E_{coh}| + E_{null})$ is used to distinguish the GW signal from the spurious glitches. The transient noise events are incoherent in phase between the detectors. They are thus detected with a lower coherent energy and a higher null energy as reconstructed detector response is inconsistent between the detectors. In the case of GW signals, we instead expect higher coherent energy and lower null energy (consistent reconstructed detector response between the detectors). As a result, GW signals have $c_c \approx 1$ and spurious events have $c_c \ll 1$. Simulations show that for low-SNR ($\lesssim 7$) events, we have $c_c \lesssim 0.7$.

The main ranking statistic used by cWB is \cite{GW150914-DETECTION}, 
\begin{equation} \label{eq:eta_c}
\eta_c = \sqrt{c_c E_{coh} K /(K-1)}.
\end{equation}

This statistic is approximately proportional to the overall network SNR.

\subsection{Wavegraph clustering algorithm}
\label{Sec:wavegraph}

In \cite{Bacon:2018zgb}, we proposed a new approach to form clusters of time-frequency pixels in model-independent searches such as cWB.

The method learns from a set of representative waveforms, what are the expected salient pixels, and how they are distributed or aligned in the time-frequency plane. The set of reference waveform can be obtained from a bank of regularly or randomly-placed templates (see e.g., \cite{Cokelaer:2007kx, Harry:2009ea}), or using an ad-hoc parameter-space discretization.

The salient pixels are extracted from the redundant time-frequency signal decompositions given by the WDM transforms using the matching-pursuit algorithm \cite{Mallat:1993:MPT:2198030.2203996}. The extracted pixels for the full set of reference signals are organized in a mathematical graph where neighbours in the set of pixels extracted from a given signal are connected by edges. Pixel neighborship is defined by a rule that tidy the pixels up in an ordered series. The graph includes the position of the selected pixels, {\it i.e.} time, frequency, and scale and their connection with other pixels.

The graph $G$ is used while analyzing the detector data with cWB at the stage where time-frequency clusters are formed. Instead of using generic geometrical clustering techniques, the selected cluster $C^*$ is obtained from the following optimization problem:
\begin{equation}
\label{eq:rankingstat}
C^{\star} = \argmax_{C \in G} \sum_{p \in C}  E[p] - \lambda \bar{E}(f_p, M_p),
\end{equation}
where $E[p]=\max_{\theta,\phi} E_{\theta,\phi}[p]$ is a proxy of the incoherent energy and $\bar{E}(f, M)$ is the median value of $E[p]$ at frequency $f$ and scale $M$ over all times $t$.

The selected cluster therefore maximizes the total incoherent energy in the cluster after removing the average contribution due to (stationary) noise. The second term in Eq.~(\ref{eq:rankingstat}) can also be viewed as a penalization, whose strength is defined by $\lambda$, that promotes smaller clusters (see \cite{Bacon:2018zgb} for more information). The selected cluster is then processed following the cWB likelihood analysis.

The wavegraph algorithm combined with cWB was tested in \cite{Bacon:2018zgb} to search for BBH signals  (total mass ranging from $10~M_{\odot}$ to $70~M_{\odot}$) in simulated Gaussian noise. Overall, this scheme shows a relative improvement of $22-26\%$ in the event rate recovery with respect to cWB alone. 

\section{Signal consistency test for wavegraph}
\label{Sec:consistency_test}

In our previous study using simulated Gaussian noise~\cite{Bacon:2018zgb}, we have shown that cWB combined with wavegraph has larger noise background than the cWB. With real detector noise, the noise background is expected to increase due to non-Gaussian noise transients in the detector.

Model-based searches use ``signal consistency tests'' to reject the event candidates identified by the matched filtering algorithm due to the noise transients. The different versions of such test include the $\chi^2$ test \cite{Allen:2004gu,Nitz:2017lco,Canton:2013joa}, bank $\chi^2$ \cite{Dhurandhar:2017aan} or $\xi^2$ tests~\cite{Messick:2016aqy}..

The $\chi^2$ signal consistency test was first introduced in \cite{Allen:2004gu} and is based on a $\chi^2$ statistics that checks the consistency of the frequency-domain amplitude profile of the candidate signal with that of the best matching template. Inspired by this principle, here we develop and validate a signal consistency test adapted to wavegraph.

\subsection{Consistency test}
\label{Sec:Cons}

The GW signal is a frequency as well as amplitude modulated signal. The wavegraph clusters carry the information of the frequency modulation in terms of the location of the TF pixels and their connection. In addition to the pixel location, the pixel also stores the amplitude modulation information in terms of the associated energy. We use this information in the consistency test to distinguish between the noisy transients and the GW transients as explained below.

We consider a cluster of pixels $C$ with amplitude for each pixel $p$ as $A[p]$ extracted from the data by wavegraph in association to a GW signal. Let us assume that we have the pixel amplitude model $w_s[p]$  normalized to unit norm i.e., $\sum_{p\in C} w_s[p]=1 $ for the (noise-free) template signal. The pixel amplitudes as observed in the noise-scaled data differ solely by an overall factor $a$ :
\begin{equation}
  \label{amp:model}
A_s[p]= a~w_s[p] \, .
\end{equation}

We propose a test based on the mean-square deviation between the observed and model amplitudes
\begin{equation}
  \label{eq:consistency}
X(a) = \frac{1}{\abs{C}} ~ \sum_{p \in C} {\left( A[p] - A_s[p] \right)^2},
\end{equation}
where $\abs{C}$ is the size of $C$.

Minimization of $X$ wrt $a$ given 
\begin{equation}\label{eq:consistency2}
X(\hat{a}) = \frac{1}{\abs{C}} ~ \sum_{p \in C} {\left( A[p] - {\hat{a}} w_s[p] \right)^2}
\end{equation}
where $\hat{a}$ minimizes $X(a)$, such that
\begin{equation} \label{Eq:estimation}
\hat{a} = \frac{\sum_{p \in C}  A[p] w_s[p]}{\sum_{p\in C} w_s^2[p]}.
\end{equation}

Unfortunately, a pixel in the graph is not necessarily associated with only one template signal. Pixels generally belong to multiple template clusters. As a result, a pixel amplitude can take a range of model amplitudes $w_s[p]$. The range of $w_s[p]$ can be characterized by its mean $\bar{w}_s[p]$ and variance $\sigma^2_s[p]$. The variance is likely to increase with the number of template clusters passing through the considered pixel. We thus use $\bar{w}_s[p]$ in place of $w_s[p]$ in the $\chi^2$ test defined in Eq.~(\ref{eq:consistency2}) and down-weight pixels with large variance in the sum defined in the next subsection. 

\subsection{Bias correction}

\begin{figure}[h]
 	\centering
 	\includegraphics[scale=0.23]{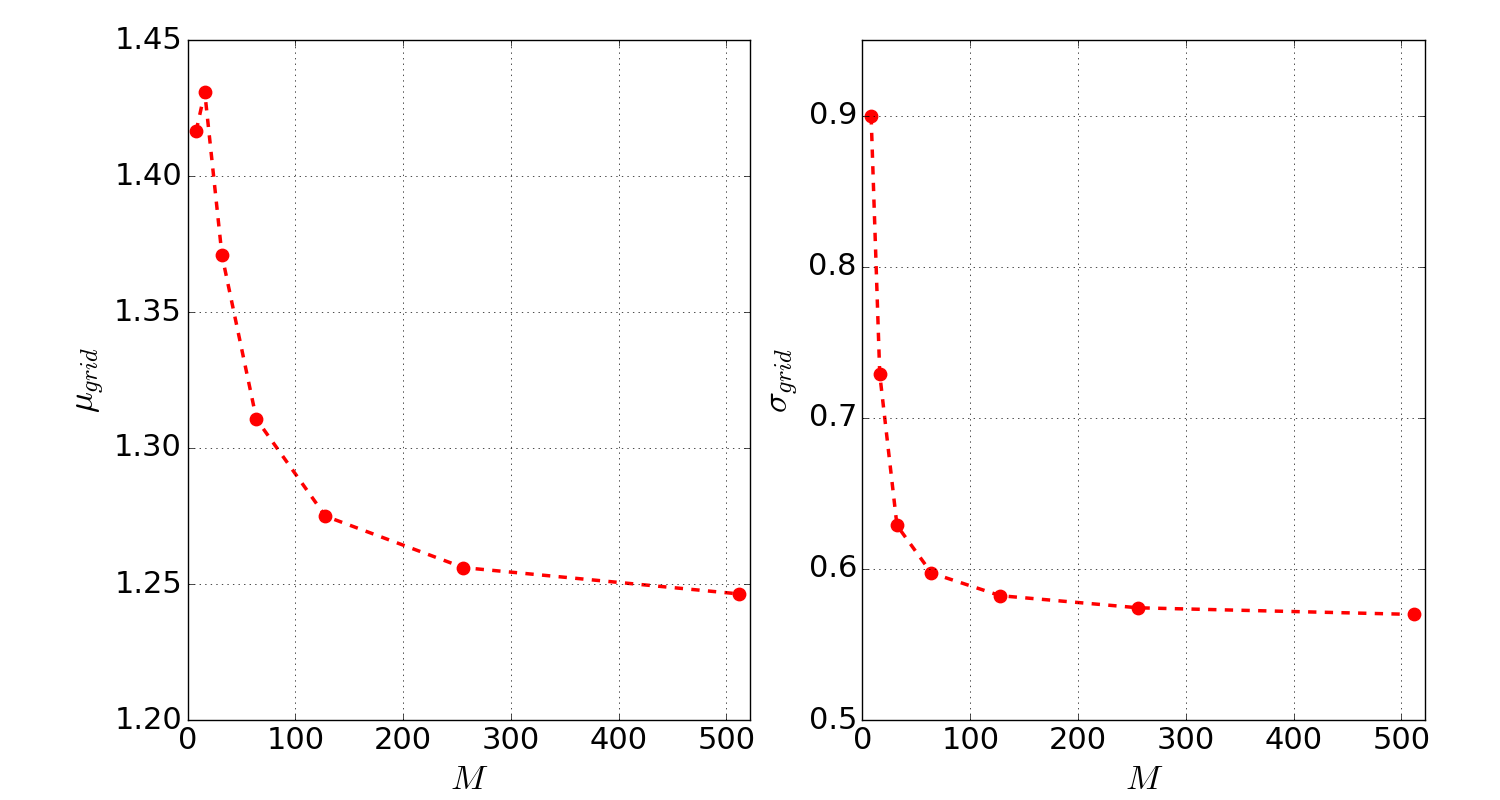}
 	\caption{Bias $\mu_{grid}$ and standard deviation $\sigma_{grid}$ of the observed pixel amplitude vs. the scale parameter $M$, estimated using LIGO O1 data.}
 	\label{fig:coarse_sky_grid}
\end{figure}

The pixel amplitude $A[p]\equiv\sqrt{E[p]}$ combines a signal term (if present) and a noise contribution. The latter creates a bias with respect to the expected value $A_s$. Since $E[p]=\max_{\theta,\phi} E_{\theta,\phi}[p]$, the pixel amplitude results from a maximization over sky positions selected on a sky grid. The result of this maximization varies depending on the duration of the Wilson-Daubechies-Meyer basis functions compared to the total span for the propagation delay over the sky. This leads to an average bias $\mu_{grid}$ and variance $\sigma^2_{grid}$ that varies with the scale parameter $M$, as shown in Figure~\ref{fig:coarse_sky_grid}. In order to produce the estimates in Figure~\ref{fig:coarse_sky_grid}, we use segments ($\sim 600$ seconds) of LIGO O1 data with reasonably stationary noise (no glitch). This simulation provides a set of estimated bias and variance at each scale. Both those quantities decrease with the scale, since, for larger scales, the propagation delay is small compared to the typical duration of the WDM functions, resulting in a smaller change of the WDM amplitudes.

The observed amplitude is corrected from the bias, and the noise variability is included to the $\chi^2$ statistics, leading to
\begin{equation}\label{eq:consitency3}
 \xi = \frac{1}{\abs{C}} ~ \sum_{p \in C} \frac{\left( A[p] - \mu_{grid}[p] - \hat{a} ~ \bar{w}_s[p] \right)^2}{\sigma^2_{grid} [p]+  \hat{a}^2\sigma_s^2[p]}. 
\end{equation}

For a GW signal with sufficient signal-to-noise ratio, we expect the statistic $\xi$ to be small, while, for the noise glitches, the large amplitude discrepancy leads to large $\xi$. We term this as a signal consistency test which is a measure of consistency of the data to the signal model. We expect that longer is the signal duration, more effective will be the test as it combine the residual power for more number of pixels.

\subsection{Validation}
\label{Sec:valid}

In this subsection, we validate the proposed consistency test $\xi$ using simulated CBC signals in real LIGO noise. We consider 16 days of data from sixth LIGO science run (S6, 2009-2010~\cite{LIGO:2012aa}) and 48.6 days of data from the first advanced LIGO observing run (O1). S6 data is particularly suited to carry out any noise rejection technique as is include a large population of noise features.

A graph is generated using a set of CBC waveforms with the SEOBNRv2 model ~\cite{SEOBNRv2,Purrer:2015tud} that spans a total mass range between $10-50~M_{\odot}$ with mass ratio up to 3. We use 4 different scales ranging in $M= 2^3$ to $2^7$ and a sampling frequency of $1024$ Hz.

The data is analyzed with {\texttt{cWB}} with wavegraph using this graph. We estimate the analysis background using the time-slide technique\footnote{The detector data are artificially shifted by nonphysical time-delays ($\gtrsim 1~s$) much larger than the physical wave propagation delay between detectors. This allows to estimate the chance probability of noise transients mimicking a GW signal in coincidence at the two detectors \cite{GW150914-DETECTION}.}. We accumulate $\sim 509$ years and $\sim 156$ years of background data from O1 and S6 data, respectively. We compute the consistency statistic $\xi$ for each cluster from this background data.

\begin{figure*}[!htp]
    	\hspace{-0.5 cm}
    	\centering
    	\begin{minipage}[b]{.45\textwidth}
    		\includegraphics[scale=0.35]{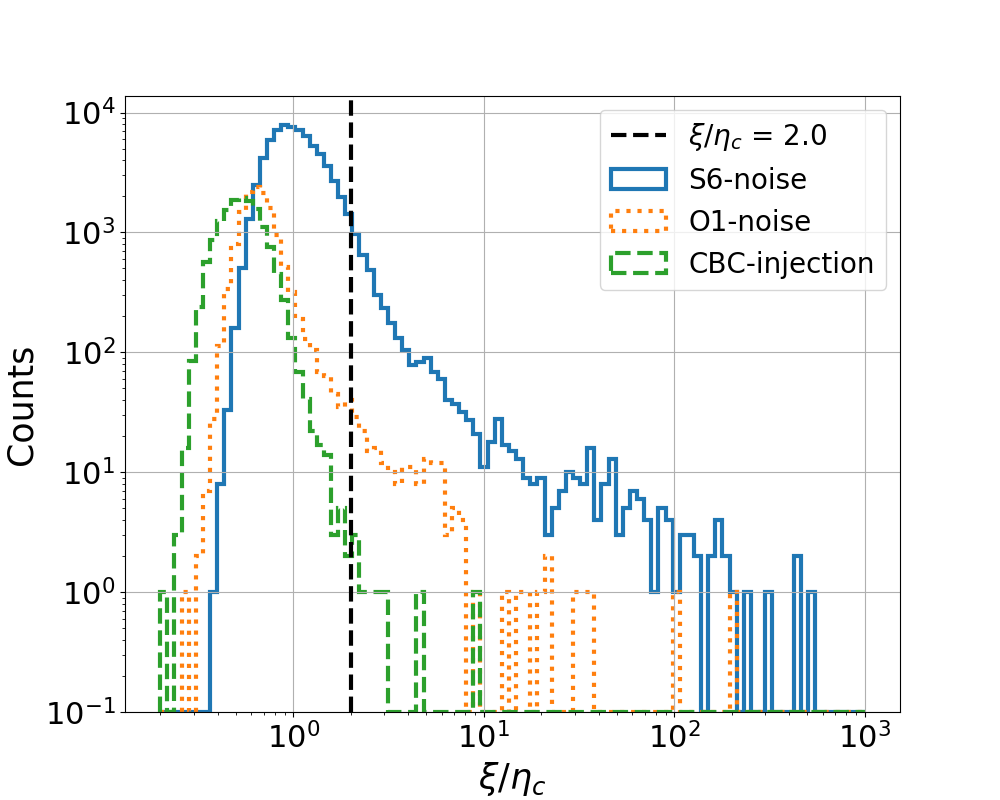}
     	\end{minipage}
    	\qquad 
    	\begin{minipage}[b]{.45\textwidth}
    		\includegraphics[scale=0.35]{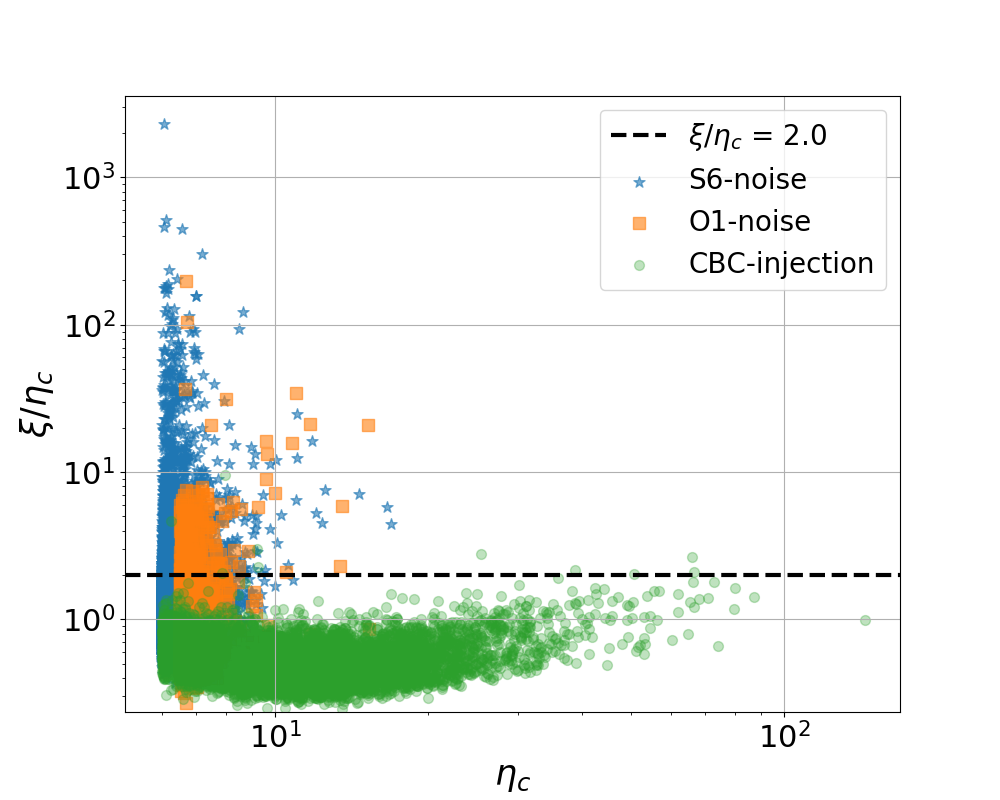}
     	\end{minipage}
    	\caption{ The $\xi/\eta_c$ distribution for selected O1 noise, S6 noise and CBC injections. The panel (a) shows the distribution of consistency statistics $\xi$ and panel (b) shows the scatter plot of $\xi/\eta_c$ vs $\eta_c$. The dashed black line in both plots shows the proposed $\xi/\eta_c$ threshold.
        } \label{fig:glitchesandinjections1}
    \end{figure*}

In Figure \ref{fig:glitchesandinjections1}, we show the results for all the noise events from the background data (O1 noise in dotted orange and S6 noise in solid blue), and we compare it to events obtained from CBC injections (green dashed). The left panel shows the distribution of $\xi/\eta_c$ and the right panel shows a scatter plot of $\xi/\eta_c$ against $\eta_c$.

Clearly, the S6 data shows a larger background than O1 data. As expected, we show that the CBC injection events have low $\xi/\eta_c$ values compared to noise events irrespective of its $\eta_c$ value. At the same time, loud noise events (high $\eta_c$) have high $\xi/\eta_c$ value. The large fraction of S6 noise events have $\xi/\eta_c$ value higher than 2. These values are much higher than that of CBC events. We can thus reject the loud glitches by setting a threshold on $\xi/\eta_c$ value. For example setting $\xi/\eta_c =2$, we reject a large fraction of the noisy glitches and lose only few CBC signals.
  

\section{Analysis}
\label{SecV}
In this section, we study the performance of the cWB combined with wavegraph and the signal consistency test introducted in Sec.~\ref{Sec:Cons}. To do so, we analyse simulated GW signals from BBH mergers in the LIGO O1 noise.
  
\subsection{Parameter space}
\label{SEC:CBC-parameter-space}

In this study, we consider three distinct regions of BBH systems, referring them to as
$R_1, R_2$ and $R_3$ for compactness as tabulated in Table \ref{table:parameters-space}
which includes columns for masses, mass ratio, number of waveforms used in the template bank
and number of nodes in the graph.

We consider BH with spins aligned with the orbital momentum and with $|\chi_{1,2}| = 0$ to $0.989$. We compute a template bank \cite{Harry:2009ea} using the \texttt{SEOBNRv2\_ROM\_DoubleSpin} waveform approximant and a minimum match \footnote{The minimum match defines the minimum overlap between an arbitrary signal in the considered range and the waveforms in the template bank. A minimum match of $0.97$ decides the density of template bank and corresponds to an overall event loss of $10\%$.} of $0.97$ for low-mass $R_1$, $R_2$ regions and $0.99$ for $R_3$ region such that we get a reasonable number of templates in each region.

\begin{figure*}[!htp] 
		\hspace{-4.5 cm}
	\begin{minipage}[b]{.28\textwidth}
			\includegraphics[scale=0.35]{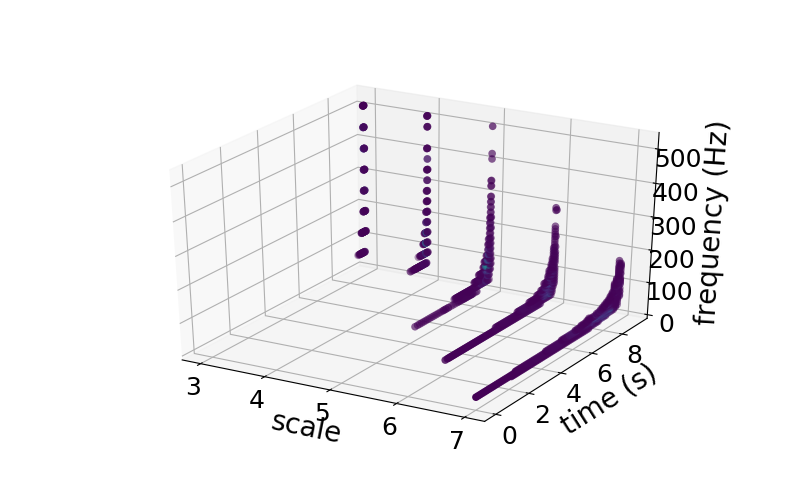}
	\end{minipage}
\hspace{1.2 cm}
		\begin{minipage}[b]{.28\textwidth}
			\includegraphics[scale=0.35]{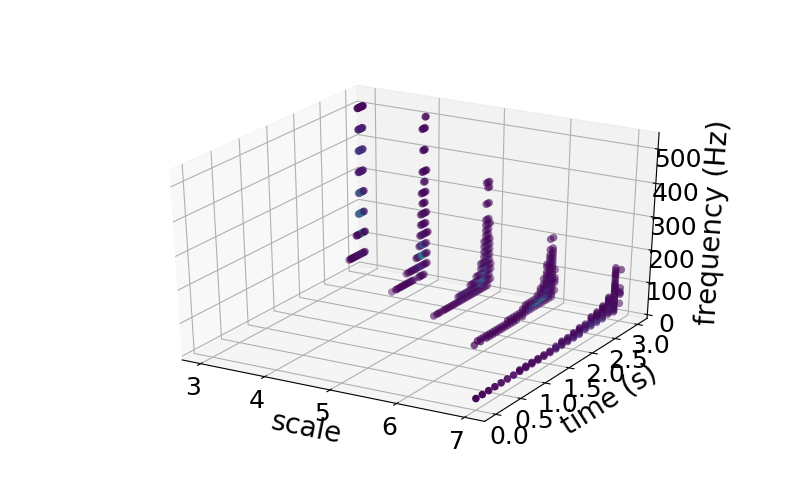}
		\end{minipage}
\hspace{1.2 cm}
		\begin{minipage}[b]{.28\textwidth}
			\includegraphics[scale=0.35]{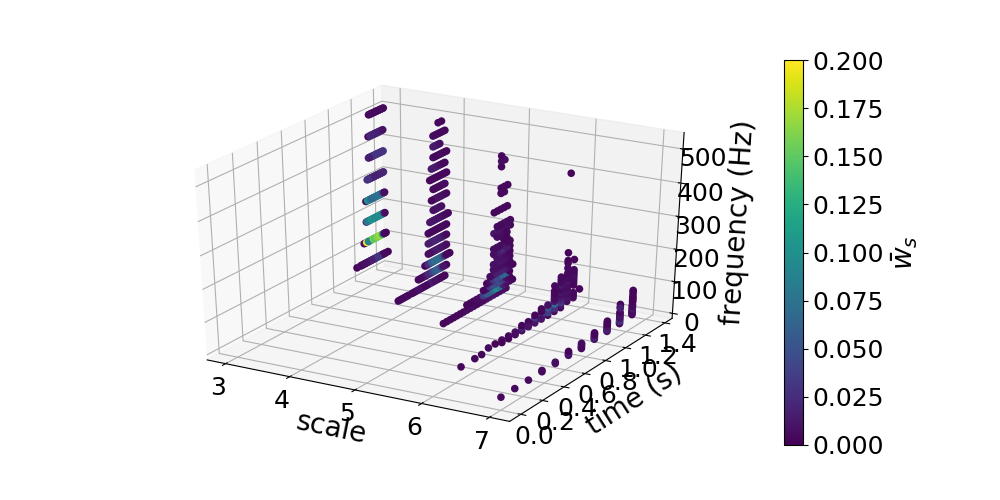}
		\end{minipage}
		
		\caption{Wavelet graph for three BBH parameter space generated using wavegraph algorithm. Left panel for $R_1$ space, middle panel for $R_2$ space and right panel for $R_3$ space. The color shows the $\bar{w}_s$ for selected wavelets.}
		\label{fig:wavegraph}
	\end{figure*}

We generate a wavelet graph for each of the $R_1$, $R_2$, and $R_3$ regions from the associated template banks using the LIGO O1 power spectral density. The time-frequency pixels in the graph are selected to get a signal recovery $>80\%$ \cite{Bacon:2018zgb}. Figure~\ref{fig:wavegraph} shows the wavelet graphs for the $R_1$ (left), $R_2$ (middle), and $R_3$ (right) parameter spaces. The figure displays the location of the selected wavelets and the colors indicate the model amplitude $\bar{w}_{s}$. The variability and range of $\bar{w}_{s}$ increases with larger masses; {\it i.e.} moving from $R_1$ to $R_3$.

The number of time-frequency pixels in the graph decreases with the total mass; following the trend of the number of templates in the bank. Their distribution also changes based on the mass. The $R_3$ graph carries more time-frequency pixels at low scales than at high scales. This picture reverses for $R_1$ region. This is primarly because high mass BBH are short duration signals which are efficiently recovered with short duration wavelets. All in all, the connectivity of the graphs is with 80 to 150 connections between the graph nodes.

\begin{table} [htb!]
		\setlength\extrarowheight{5pt}
		\begin{center}
			\begin{tabular}{ |p{1 cm} | p{2 cm} | p{ 1. cm} | p{1.5 cm} | p{2 cm} |}
				\hline
				\text{BBH}  & \text{ $m_1, m_2$}& \text{  ${q}$} & \text{No. of }  & \text{No. of}     \\
				\text{region}  & \text{in $~~M_\odot~~$  }&   &  \text{templates}&\text{wavelets}  \\ \hline 
			$	R_1$  &  $5.0$ to $12.5$  &  $<3$ & 11829 & 1333 \\ \hline
				$R_2$  &  $12.5$ to $20.0$ & $<2$   & 2546 & 700 \\ \hline
				$R_3$ &  $20.0$ to $35.0$ & $<2$ & 637 & 619   \\ \hline
			\end{tabular}
		\end{center}
		\caption{Details of three distinct regions in BBH parameter space. We tabulate the component masses $m_1$ and $m_2$, the mass ratio $q$, number of templates and the number of wavelets in the constructed graph.}
		\label{table:parameters-space}
\end{table} 

\subsection{Estimation of the noise background}

\begin{figure}[!htp]
    	\centering
\includegraphics[scale=0.35]{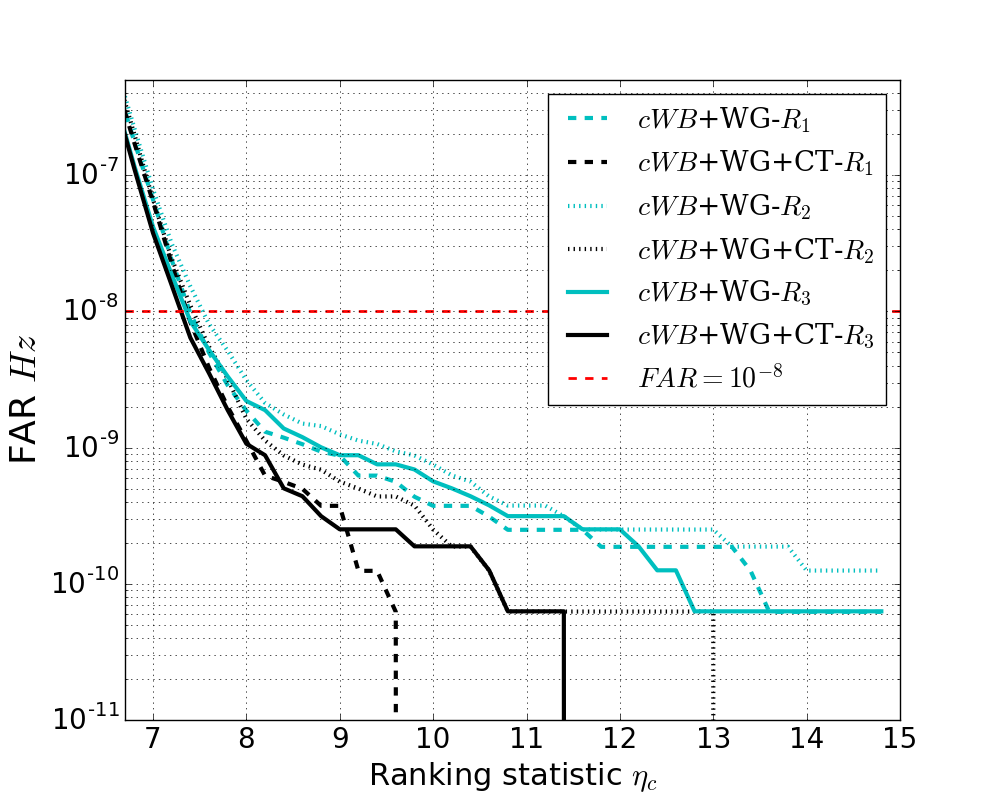}
\caption{Background False Alarm Rate (FAR) {\it vs} the statistic $\eta_c$. No noise vetoes has been applied here, besides the signal consistency test described in Sec.~\ref{Sec:consistency_test}. The solid, dotted and dashed curves correspond to $R_3, R_2$ and $R_1$ regions respectively. The lines in cyan correspond to {\it cWB} with wavegraph (WG) while the lines in black corresponds to {\it cWB} with WG and consistency test (CT). The horizontal dashed-red line shows the reference FAR level of $10^{-8}$~Hz (0.3 event per yr).}
\label{fig:bkg_consistency_plot}
\end{figure}

\begin{figure}[h!]
	\centering
	\includegraphics[scale=0.35]{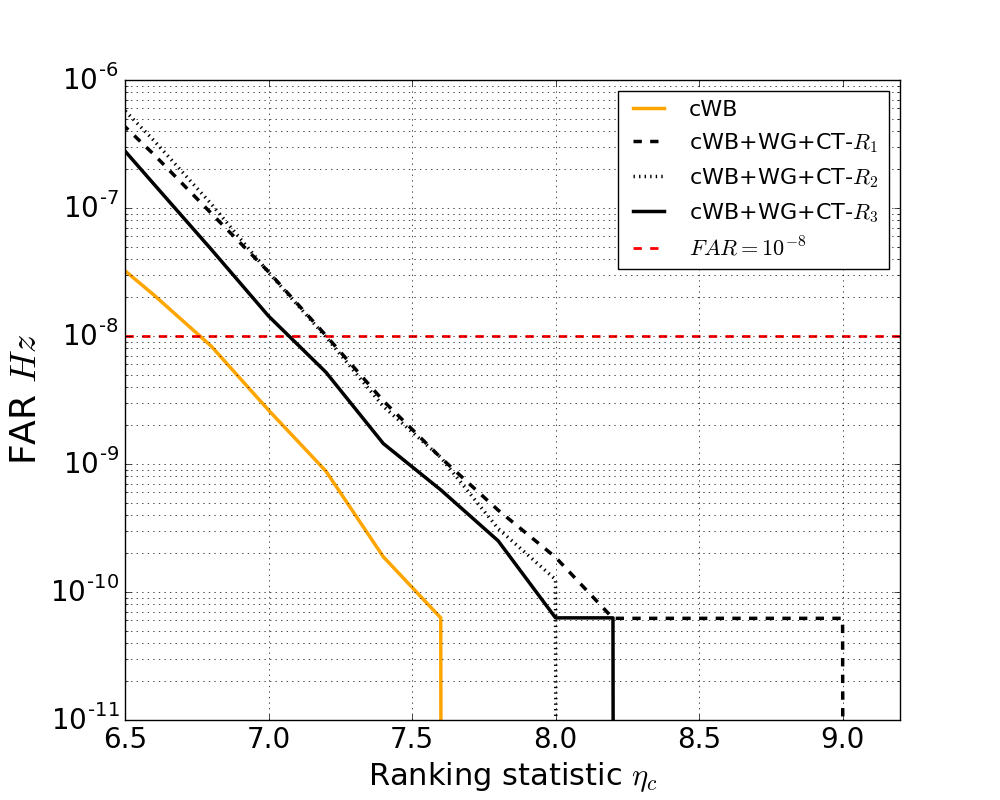}
	\caption{Background False Alarm Rate (FAR) {\it vs} the statistic $\eta_c$. Noise vetoes discussed in Appendix~\ref{cwb-veto} have been applied. The solid, dotted and dashed curves correspond to $R_3, R_2$ and $R_1$ regions. The orange (resp. black) color is for cWB (resp. cWB with wavegraph and consistency). The horizontal dashed-red line indicates the reference FAR $=10^{-8}$~Hz  (0.3 event per yr).}
\label{fig:bkg_plot}
\end{figure}

A critical part of any GW search algorithm is to measure the distribution of accidental triggers due to the noise of the detector. We estimate the false alarm rate (FAR), {\it i.e.} the rate of accidental triggers, for the coherent WaveBurst algorithm with and without wavegraph using LIGO O1 noise.

The cWB algorithm includes a series of vetoes to reject noise transients in the data. Those vetoes are discussed in details in Appendix~\ref{cwb-veto}. The noise background curve in terms of FAR vs $\eta_c$ for cWB combined with wavegraph are shown in Figure~\ref{fig:bkg_consistency_plot} without applying the noise vetoes and in Figure~\ref{fig:bkg_plot} with the noise vetoes. We compare the noise backgrounds of cWB and cWB with wavegraph using the three graphs presented earlier (dashed, dotted and solid curves are for the $R_1$, $R_2$, and $R_3$ graphs) and the signal consistency test described in Sec.~\ref{Sec:consistency_test} accepting all the events below $\xi/\eta_c =2$.

In Figure~\ref{fig:bkg_consistency_plot} the cyan (black) curves are for {\it cWB} with wavegraph without (with) signal consistency test (labelled CT). We observe that the long tails of the noise background curve (cyan) for {\it cWB} with wavegraph get trimmed after applying the signal consistency condition of $\xi/\eta_c =2$ (black). In the $R_1$ region, we are able to remove more glitches. They can be disentangled more efficiently from the signal as the latter has a longer duration. 



In Figure~\ref{fig:bkg_plot} the solid orange, solid black, dotted black and dashed black lines are for cWB, cWB including wavegraph and the consistency test with $R_3$, $R_2$ and $R_1$ respectively. From this calculation, we compute the threshold on $\eta_c$ associated to the reference FAR value of $10^{-8}$ Hz (0.3 events per yr). We obtain $\eta_c =$ 6.76, 7.07, 7.19 and 7.2 for cWB, cWB including wavegraph and consistency test for $R_3$, $R_2$ and $R_1$ respectively.

It appears that cWB including wavegraph and consistency test has a larger background compared to cWB. Nevertheless, as we will see in the next section, owing to the better recovery of the correlation coefficient, this algorithm shows an improved overall sensitivity.

\subsection{Simulations}

We now estimate the sensitivity of cWB with wavegraph algorithm using simulated BBH signals in the LIGO O1 noise covering  the $R_1$, $R_2$ and $R_3$ parameters space as defined in Table~\ref{table:parameters-space}. Injections are generated using the \texttt{SEOBNRv2} waveform model and are uniformly distributed over the binary parameter space ($m_1$, $m_2$, spins, sky-location, $\cos \iota$), and distributed over uniform volume up to a fixed maximum distance: $R1$ is up to $1$~Gpc, $R2$ is up to $1.5$~Gpc and $R3$ is up to $3$~Gpc.

\begin{figure}[h!]
 		\centering
 		\includegraphics[scale=0.27]{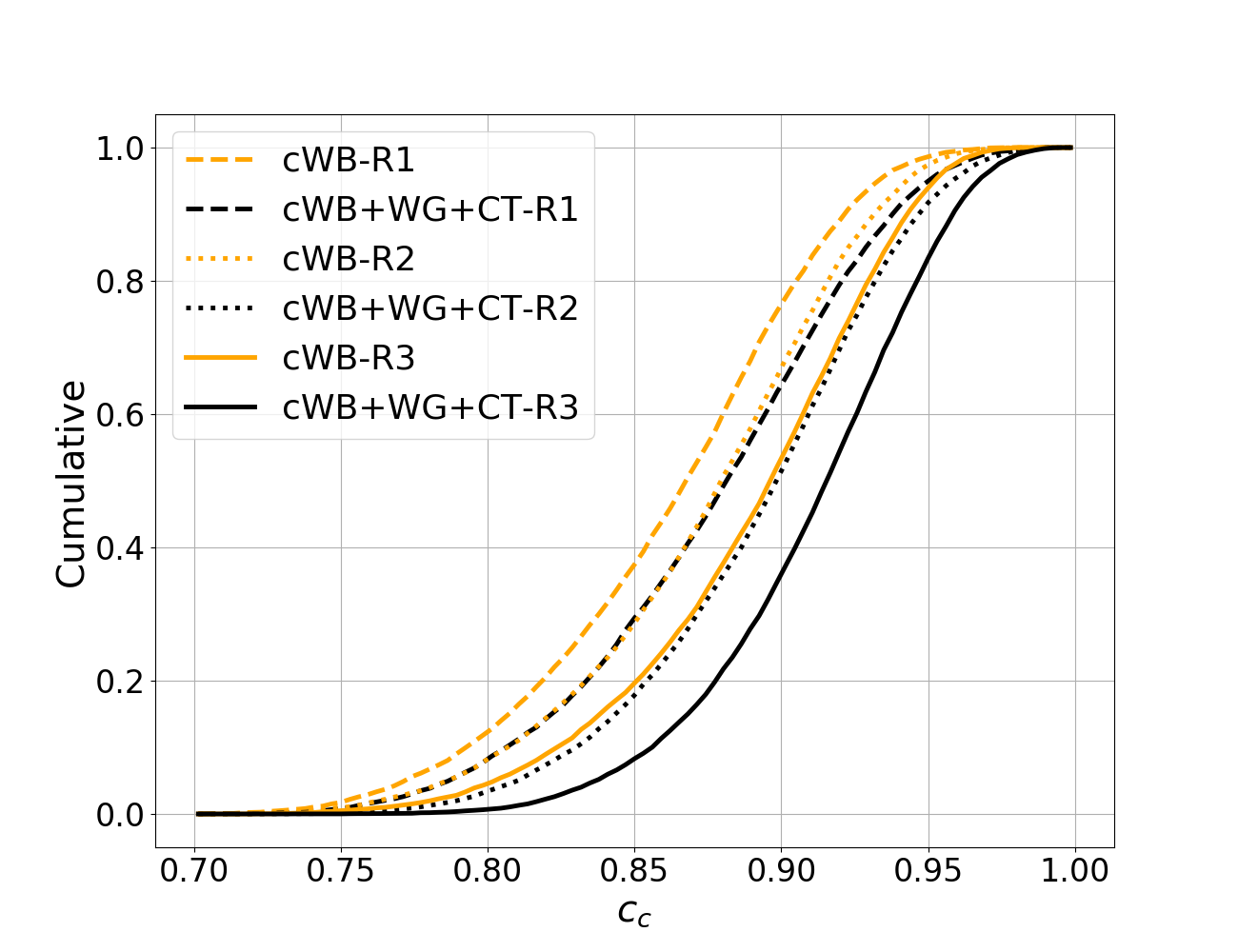}
 		\caption{Cumulative distribution of the network correlation for the recovered events. The orange (black) line is for cWB (cWB+WG+CT). The solid, dotted and dashed lines are for the $R_3$, $R_2$ and $R_1$ regions respectively.}
                \label{fig:netcc}
 	\end{figure}

Figure~\ref{fig:netcc} shows the cumulative distribution of the network correlation value for recovered events at given FAR threshold of $10^{-8}$~Hz. The orange (resp. black) curve is for cWB (resp. cWB including wavegraph and consistency test). The solid, dotted and dashed lines are for the $R_3$, $R_2$ and $R_1$ regions respectively.

We observe that cWB with wavegraph recovers injections with higher correlation coefficient $c_c$. This effect is more pronounced for the high-mass region. As a result, the events recovered by cWB with wavegraph have higher network correlation than the events recovered by cWB alone. That is primarily due to the wavegraph clustering method which helps to collect more coherent pixels as well as the consistency test which rejects inconsistent noisy triggers. This feature also plays an important role in removing the noisy events from the analysis background.

\begin{figure}[H]
	\centering
	\begin{subfigure}{0.48\textwidth}
		\includegraphics[scale=0.27]{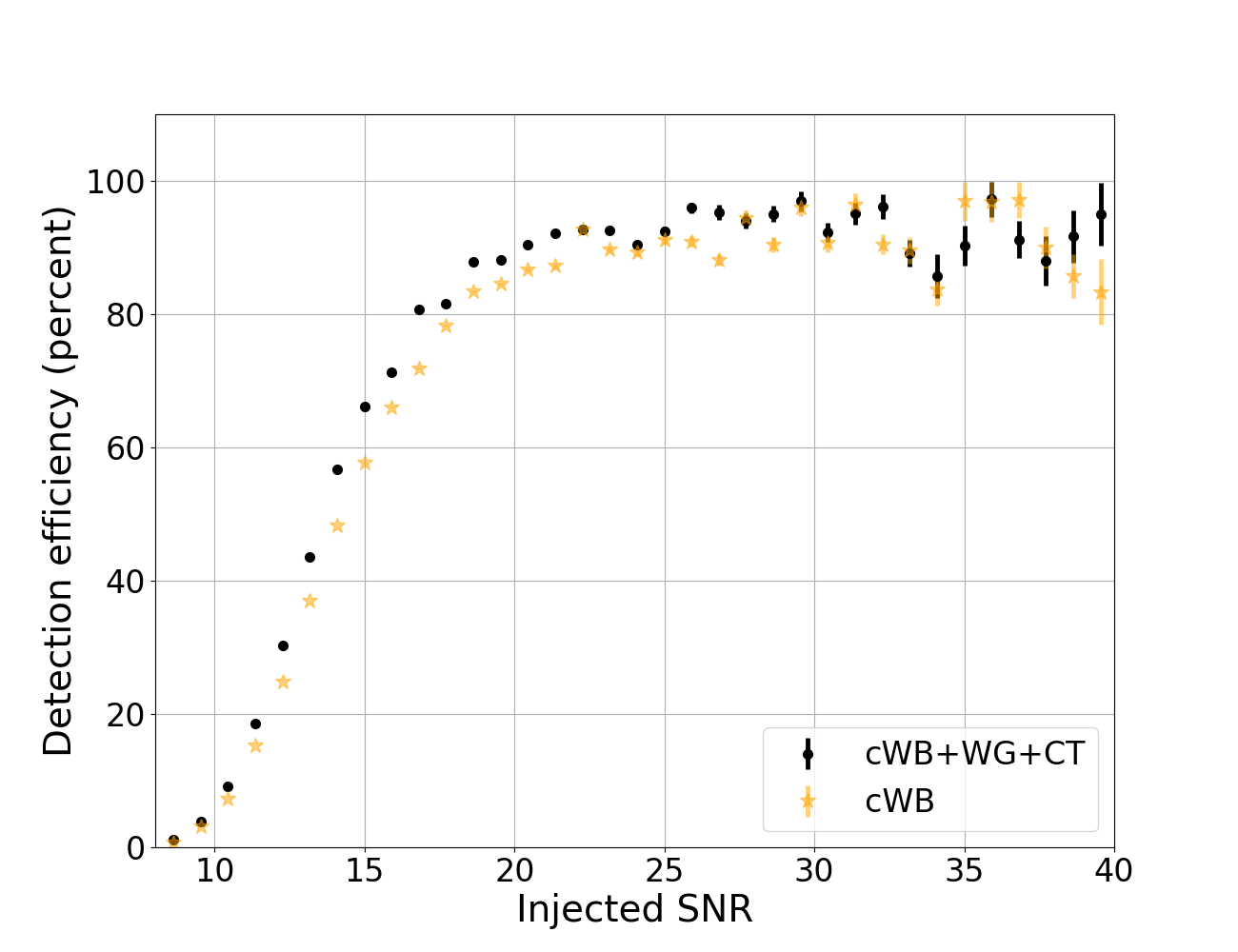}
		\caption{R1 region} \label{fig:efficiency_hist_final}
	\end{subfigure}
	\medskip
	\begin{subfigure}{0.48\textwidth}
		\includegraphics[scale=0.27]{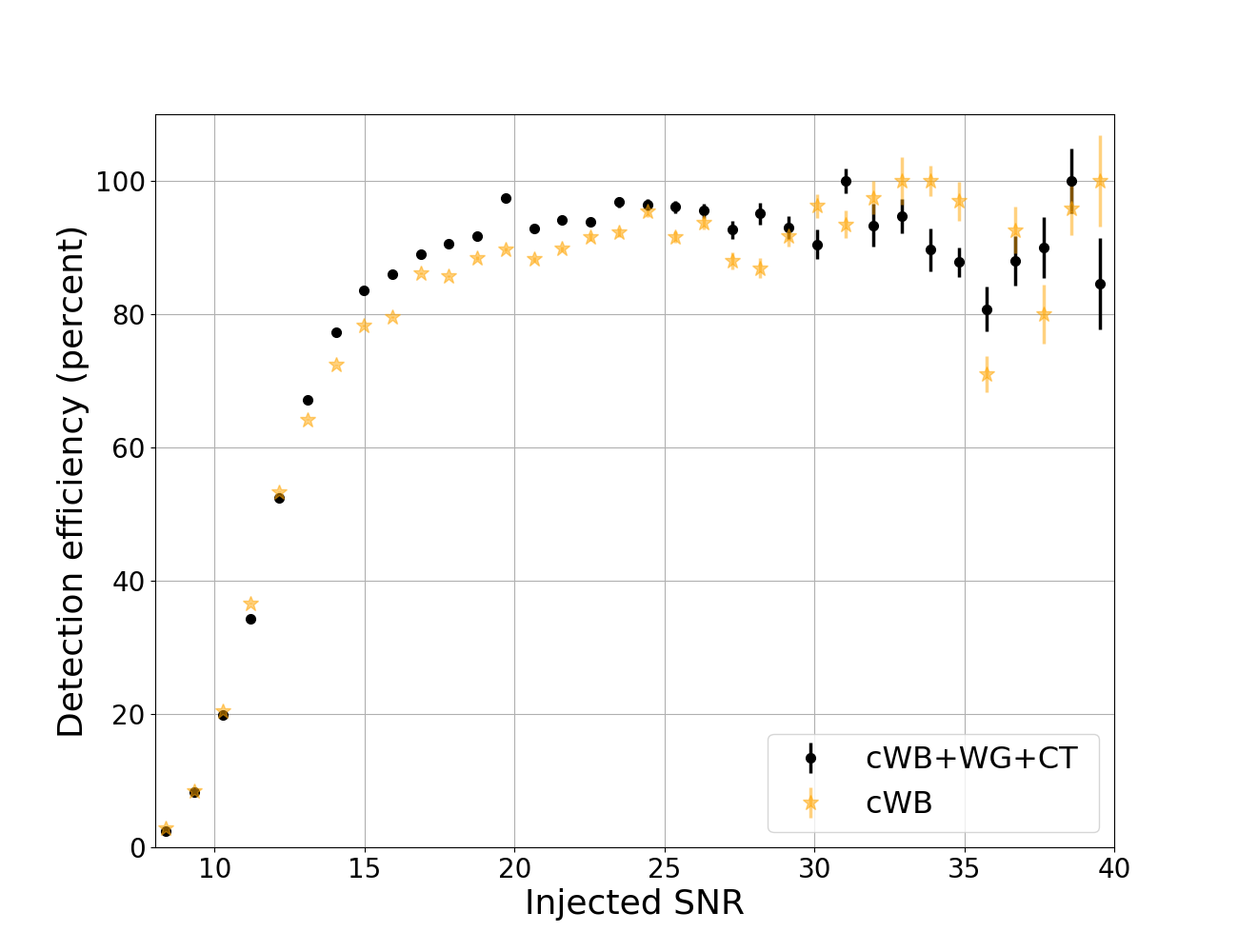}
		\caption{R2 region} \label{fig:efficiency_hist-R2_final}
	\end{subfigure}
	\medskip
	\begin{subfigure}{0.48\textwidth}
		\includegraphics[scale=0.27]{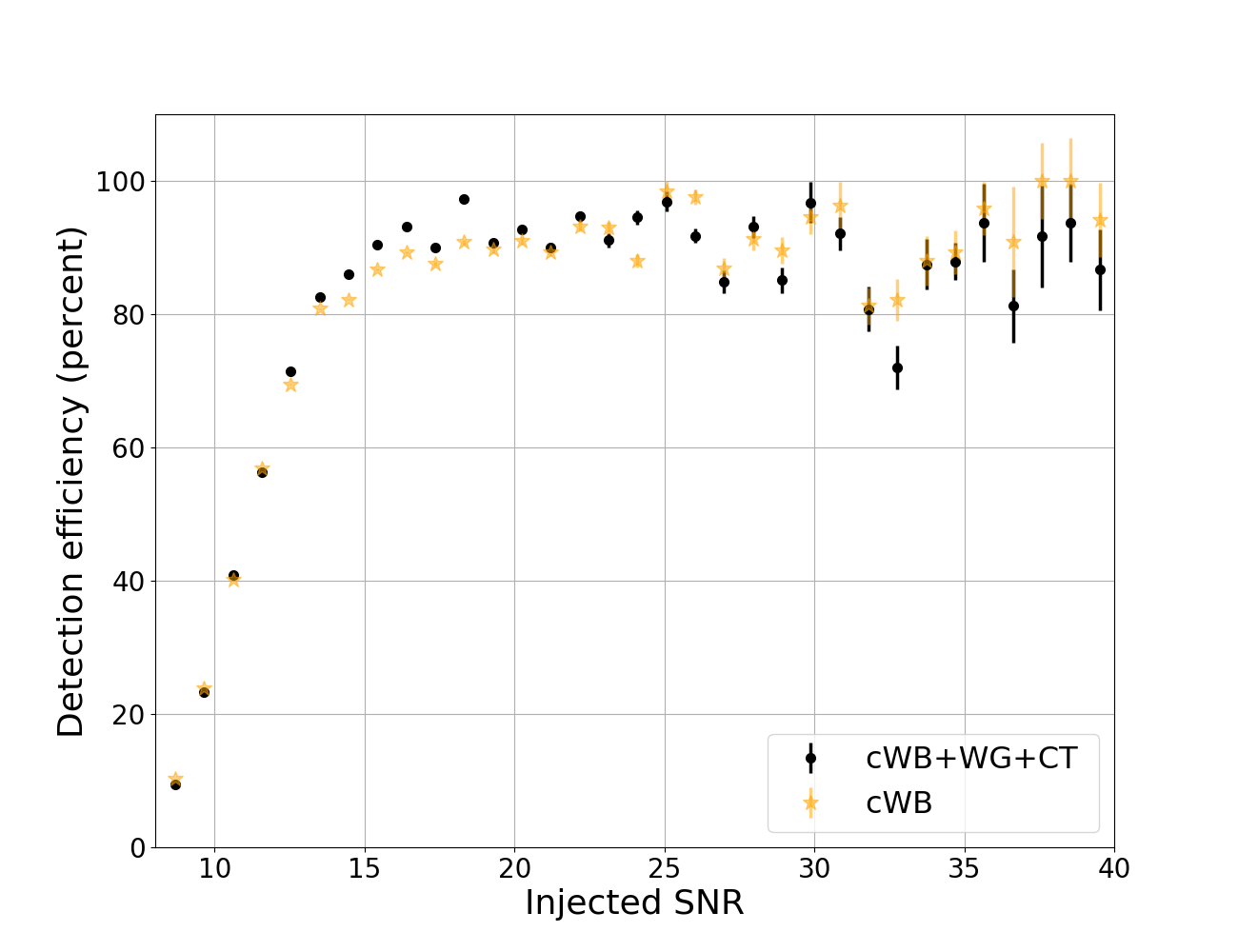}
		\caption{R3 region} \label{fig:efficiency_hist-R3_final}
	\end{subfigure}	
	\caption{Detection efficiency (in percent) vs the injected network
		SNR given for cWB (orange stars) and for cWB with wavegraph
		(black circles) with 1-sigma error bars. The (a), (b) and (c) panels 
		correspond to the $R_1$, $R_2$ and $R_3$ simulations respectively.}
        \label{fig:efficiencyhist}
\end{figure}

Figure~\ref{fig:efficiencyhist} shows the detection efficiency curve for cWB as well as cWB with wavegraph algorithm in three different BBH parameter space. cWB with wavegraph and the consistency test recovers slightly more events in the low-mass range than in the high mass range compared to cWB. However, what is more important is that due to different clustering methods, amongst all the recovered events $62\%$, $83\%$ and $85 \%$ of the events are common in $R_1, R_2$ and $R_3$ regions respectively. With cWB including wavegraph and consistency test, we recover additional $38\%$, $17\%$ and $15 \%$ events from $R_1, R_2$ and $R_3$ regions respectively.
        
	\begin{figure}[h!]
 		\centering
 		\includegraphics[scale=0.27]{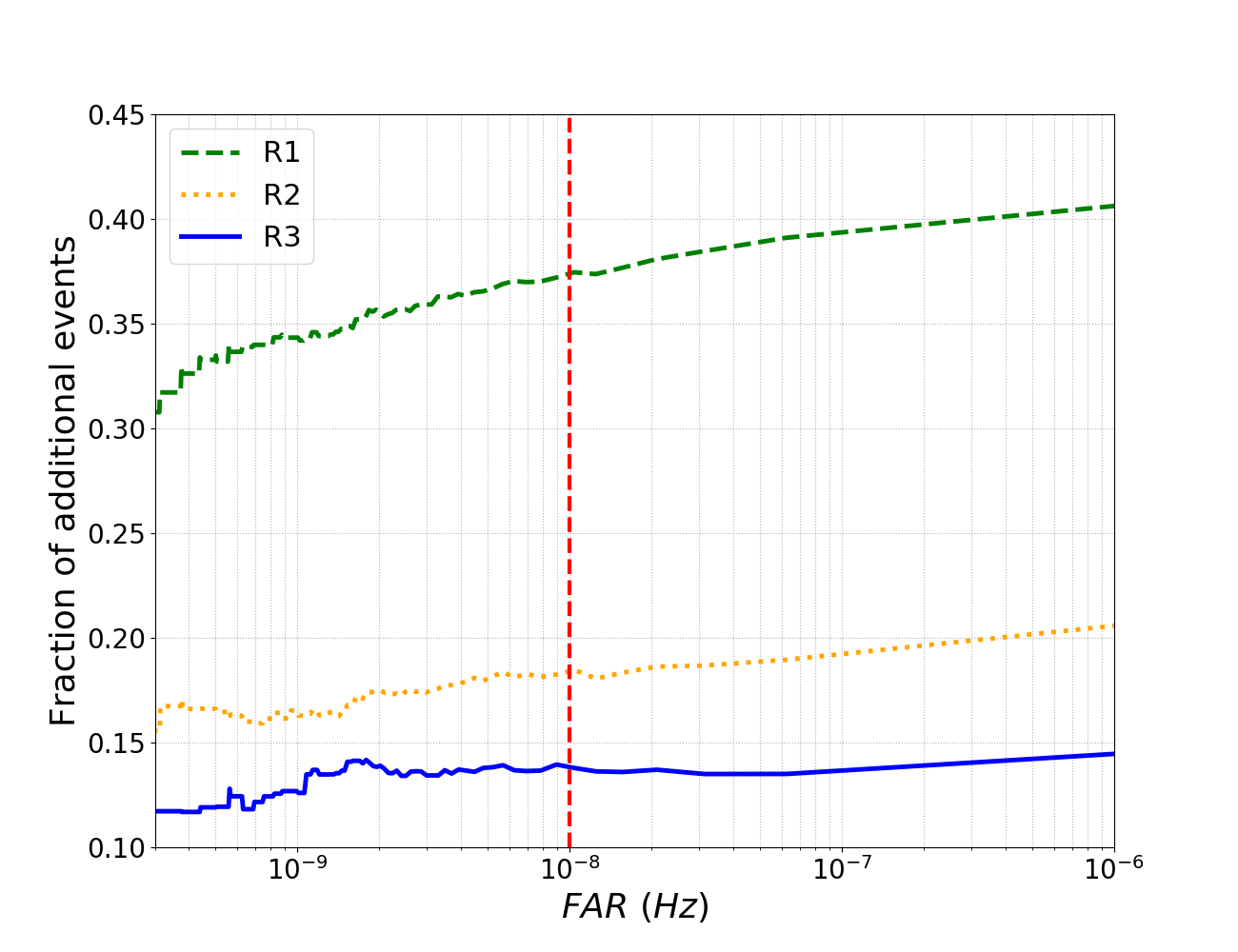}
 		\caption{Fraction of extra detected events by the combined pipeline cWB+WG+CT compared to cWB alone for the $R_1$ (green-dashed), $R_2$ (orange-dotted) and $R_3$ (blue-solid) regions as a function of the FAR threshold (Hz).}
                \label{fig:ratioevents}
 	\end{figure}

Figure~\ref{fig:ratioevents} provides a broader view of the detectability improvement. We show the fraction of additional detections (i.e., ratio of number of additional events detected by cWB+WG+CT -- but missed by cWB -- divided by number of events recovered by cWB) for a range of FAR thresholds. The relative increase of the detection rate appears to be robust and approximately constant for the considered range of FAR (note that the estimate is less reliable at lower FAR due to lower statistics). For FAR$=10^{-8}$ Hz, we retrieve the relative gains stated above.

\section{Conclusion}
\label{SecVI}

We have shown in \cite{Bacon:2018zgb} that the sensitivity of cWB search can be improved by incorporating the astrophysical information at the clustering stage. The clustering uses a mathematical graph to store the astrophysical information in the form of TF pixels and their connection. We tested this method in the cWB algorithm for the GW signals emitted by BBH systems with simulated Gaussian noise using advanced two LIGO detectors and Virgo detector network. 

In this work, we have proposed a new signal consistency test $\xi$ for cWB with wavegraph algorithm.  This test uses the amplitude profile information to distinguish between the GW transients from the noisy glitches.  For real GW signals, the consistency test value $\xi$ is small  compared to the noisy glitches. We use this information to veto noisy glitches.  We applied and tested cWB with wavegraph along with signal consistency test for simulated BBH GW signals in O1 data. We test this algorithm in three different non-precessing BBH parameter space. We observe that using cWB including the wavegraph as well as signal consistency test, we remove a large fraction of loud glitches in O1 data.

   
The cWB and cWB including wavegraph and consistency test recovers similar fraction of the events however, out of those $0.62$, $0.83$, $0.85$ fraction of events
are common in $R_1$, $R_2$ and $R_3$ regions. This clearly implies that the union of both bring in $0.38$, $0.17$, $0.15$ fraction of additional events in $R_1$, $R_2$ and $R_3$ region respectively.


    
    In summary, the wavegraph clustering approach along with the consistency test is general enough that it can be extended to a broad range of astrophysical systems which emit long duration GW signals in the ground based detectors. We are exploring this approach for detection of GW from long duration bursts signals such as eccentric binary black holes, accreting BH systems etc.

\section{Acknowledgements}
This research has made use of data \cite{Vallisneri:2014vxa}, software and/or web tools obtained from the Gravitational Wave Open Science Center (\url{https://www.gw-openscience.org}), a service of LIGO Laboratory, the LIGO Scientific Collaboration and the Virgo Collaboration. LIGO is funded by the U.S. National Science Foundation. Virgo is funded by the French Centre National de Recherche Scientifique (CNRS), the Italian Istituto Nazionale della Fisica Nucleare (INFN) and the Dutch Nikhef, with contributions by Polish and Hungarian institutes.

This research was supported by the CEFIPRA grant No IFC/5404, by the
European Union's Horizon 2020 research and innovation programme under
grant agreement No 653477, by the French Agence Nationale de la
Recherche (ANR) under reference ANR-15-CE23-0016 (Wavegraph project),
and by CNRS PICS/Inde. VG acknowledges Inspire division, DST,
Government of India for the fellowship support. VG would like to thank
IISER-TVM for providing facility to complete the initial part of this
work. The
authors are grateful to the LIGO Scientific Collaboration and Virgo
Collaboration for giving access to the simulation software used here,
and specifically to the cWB team for their help and support. The
authors are grateful to Sergey Klimenko, and Marco Drago for their valuable comments and
suggestions.


\appendix
\section{Noise veto methods for coherent WaveBurst algorithm}
\label{cwb-veto}

For completeness, we review the glitch rejection methods included in the coherent WaveBurst algorithm and used in the context of transient searches in the LIGO/Virgo O1 and O2 data sets.

\subsection{Norm} 
\label{section:norm}

The frequency of CBC chirp signal evolves with time. The chirping signal energy is distributed over multiple TF maps and each part of the signal is captured by a specific wavelet. The signal is represented by the collection of wavelets from different scales/levels. The frequency change observed for noise glitches differs from the GW signal. Most of the loud glitches are localized in the narrow frequency range and thus represented by a collection of wavelets with few levels. cWB uses this information to separate glitches from astrophysical CBC signal. The $E_{ft}$ is the time-frequency energy of the event in wavelet domain and $E_t$ is the reconstructed event energy in time domain.

The average number of WDM resolutions used for event reconstruction is estimated by the oversampling factor $N_{norm}= E_{ft}/E_t$. The $N_{norm}$ value strongly depends on the event SNR. The low SNR events are typically reconstructed with $N_{norm} \sim 1$ whereas very high SNR events are reconstructed with $N_{norm}$ equal to two times of the number of WDM resolution used for the analysis.

\subsection{Noise statistics, $\zeta$}
\label{section:chi}

For real GW signal, the noise energy in the data after subtracting reconstructed signal from the strain data is low. For the noisy glitch, high residual noise energy remains in the data. The cWB algorithm uses this information to discriminate the real GW event from the noisy glitches. The residual noise energy in the data $E_n$ is estimated as sum of residual noise energy $E_r$ and energy contribution from Gaussian noise remaining in the data $E_g$. The noise statistics $\zeta^2 = E_n/N_{Dof} =(E_r+E_g)/N_{Dof}$. Here, $N_{Dof}$ is the number of degrees of freedom give by number of detector times number of pixels. The $\zeta^2$ is estimated in both time and time-frequency domain. The GW signals have a $\zeta^2$ value below 1. A significant deviation from unity is an indication of presence of a glitch. By applying a threshold on $\zeta^2$ value, cWB separates the glitch from the real GW signal.

\subsection{Qveto}
\label{section:qveto}

``Blip glitches''\cite{Abbott:2016jsd} are a type of short duration noisy glitches in LIGO data of unknown origin and show rain-drop like structure in TF plane. The blip glitch in the TF plane is well-localized and maximum energy of the glitch event is derived from a small number of pixels. However, in case of CBC events, the energy is distributed over all the pixels. cWB has developed Qveto for CBC events that estimates the energy distribution of the event over different time segments.

Let $A_{max}$ be the absolute maximum amplitude of the reconstructed waveform of the event,  and $A_{1} ~\& ~A_{2} $ are the first adjacent peak amplitudes of $A_{max}$, where subscript $1$ and $2$ indicate left and right peaks respectively.  Using these values, we estimate the energy fraction of the event as $E_1= A_{max}^2+A_1^2+A_2^2$.  At the same time, we estimate the energy of the peaks (other than the mentioned three peaks) that have an amplitude greater than a fraction of 
$A_{max}$ (the search fixes this fraction value, and it also avoids counting Gaussian noise peaks), and it is denoted as $E_2$.   
In the case of a localized glitch, most of the event energy is accumulated around $A_{max}$ peak compared to other regions. But in the case of CBC signals, the energy is distributed over the full signal.  The ratio of these energies $E_2/E_1$ gives a statistic (named $Q_{veto}$) that will help to distinguish the localized glitch from CBC signals. The localized glitches will have low Qveto value as compared to CBC signals.

\subsection{Lveto}
\label{section:lveto}

Detector data contains a large population of narrow band glitches (line structures in the TF map) \textit{e.g.}, power line glitches and their harmonics. The cWB algorithm uses the statistics that computes the energy contribution to the event from the narrow frequency band.

The narrow frequency range is estimated from the loudest time-frequency pixel from the event cluster and this range is defined as $f_0-df$ to $f_0+df$ where $f_0$ is the center frequency of loud pixel and $df$ is width of the loud pixel. The cWB algorithm stores frequency mean, RMS frequency, and ratio of energy in narrow band frequency range and full energy of the event. For the narrow glitches, the energy ratio is close to 1, that means most of energy comes from narrow frequency range. These glitches are rejected by applying thresholds on these values. 

\subsection{Chirp cut}
\label{section:chirp}

This veto verifies whether the alignment of selected set of time-frequency pixels is roughly consistent with the expected model obtained from the zeroth- or
Newtonian order approximation of the CBC signal frequency evolution, namely, 
$\frac{96}{5}\pi^{8/3}\left(\frac{G \cal{M}}{c^3}\right)^{5/3} \; t\;+ \frac{3}{8} f^{-8/3} + C=0$,  where $\cal{M}$ is the (unknown) chirp mass, $G$ is the gravitational constant, $c$ is the speed of light and $C$ is a constant related to the merger time. The chirp mass parameter is estimated and error bars are produced using bootstrapping procedure using different subsets of selected pixels \cite{Tiwari:2015bda}.

The typical conditions for accepting the event as a signal event is with $N_{norm}$ greater that 2.5, $\chi^2$ less than 0.2, $Q_{veto}$ greater than  0.3 and $\cal{M}$ greater than 1.

\bibliography{reference1}

\end{document}